\documentclass[10pt]{article}

\usepackage{amsmath}
\usepackage{amssymb}
\usepackage{amsfonts}

\usepackage{graphicx}

\usepackage{cite}

\usepackage{color}

\usepackage[labelfont=bf,labelsep=period,justification=raggedright]{caption}

\makeatletter
\renewcommand{\@biblabel}[1]{\quad#1.}
\makeatother

\date{}

\begin{document}

\begin{flushleft}
{\Large
\textbf{Individual focus and knowledge contribution}
}

Lada A. Adamic$^{1,\ast}$, 
Xiao Wei$^{1}$, 
Jiang Yang$^{1}$,
Sean Gerrish$^{2}$,
Kevin K. Nam$^{1}$,
Gavin S. Clarkson$^{3}$
\\
\bf{1} School of Information, University of Michigan, Ann Arbor, MI, USA
\\
\bf{2} Dept. of Computer Science, Princeton, NJ, USA
\\
\bf{3} University of Houston Law Center, Houston, TX, USA
\\
$\ast$ E-mail: ladamic@umich.edu
\end{flushleft}

\section*{Abstract}
Before contributing new knowledge, individuals must attain requisite
background knowledge or skills through schooling, training, practice,
and experience. Given limited time, individuals often choose either to
focus on few areas, where they build deep expertise, or to delve less
deeply and distribute their attention and efforts across several
areas. In this paper we measure the relationship between the
narrowness of focus and the quality of contribution across a range of
both traditional and recent knowledge sharing media, including
scholarly articles, patents, Wikipedia, and online question and answer
forums. Across all systems, we observe a small but significant
positive correlation between focus and quality.

\section*{Introduction}
Prior to making novel or useful contributions to scholarly work, an
individual must acquire pertinent knowledge. Since acquisition of such
knowledge often involves time-consuming study or extensive experience,
most individuals tend to build their expertise in a small range of
fields; and experiencing success in one subject area may, through
positive reinforcement and the ability to obtain more resources,
result in additional focus on that subject. On the other hand, there
are myriad successful individuals who dabble in many sciences,
inventors who invent diverse gadgets, and Wikipedia editors who edit
pages on seemingly disparate topics. One might argue that their
versatility is a reflection of superior abilities that can yield greater opportunities for 
varied collaborations and cross-polination of ideas. In this paper, we present the first comprehensive, large-scale analysis of this relationship between
individual focus and performance across a broad range of
both traditional and modern knowledge collections.

Much previous research in this area has aimed to quantify the
  benefit of interdisciplinarity among researchers at the \emph{group}
  level. A study of scholarly articles in the UK, for example, found
that research articles whose coauthors are in different departments at the same
university receive more citations than those authored in a single
department, and those authored by individuals across different
universities yield even more citations on average~\cite{katz1997mcw}.
Multi-university collaborations that include a top tier-university
were found to produce the highest-impact research articles~\cite{jones2008mur}.
It has also been demonstrated that scholarly work covering a range of
fields -- and patents generated by larger teams of coauthors--tend to
have greater impact over time~\cite{wuchty2007idt}. Collaborations
between experienced researchers who have not previously collaborated
fare better than repeat collaborations~\cite{guimera2005team}. In the
area of nanotechnology authors who have a diverse set of collaborators
tend to write articles that have higher
impact~\cite{rafols08diversity}. Finally, diverse groups can,
depending on the type of task, outperform individual experts or even
groups of experts~\cite{page2007dpd}.

All of this work is evidence of a benefit in bringing together diverse
individuals.  It does not demonstrate, however, whether diversity in
research focus is beneficial at the \emph{individual} level.  One
exception is a study of political forecasting that established that ``foxes'', individuals who know many little things, tend to make better
predictions about future outcomes than ``hedgehogs'' who focus on one big thing ~\cite{tetlock2005expert}. Our work addresses
knowledge contribution in a much broader context than forecasting and,
more importantly, quantifies the relationship between individuals'
narrowness of focus and the corresponding quality of their contributions.

\section*{Results}
In order to cover a broad range of knowledge-generating activities, we study several collections of traditional scholarly publications in
addition to recent, Web-based media collections.  The traditional
media we consider include patents and research articles.  Our patent collection
consists of 5.5 million patents filed with the USPTO between 1976 and
2006.  We consider two sources of research articles:
JSTOR and the American Physical Society (APS). The JSTOR corpus consists of 2 million articles from 1108
journals in the natural sciences, social sciences, and humanities, and
the APS dataset we consider covers over 200,000 research articles in
the single discipline of physics. We complement data from these traditional publication venues with data from two recent, online types of knowledge-sharing activity: Wikipedia and a collection of
question-answering forums. Wikipedia is a collaborative online effort
to document all of human knowledge in a systematic way into a popular
internet-based encyclopedia. The Question and Answer forums we study are
Yahoo Answers (English)~\cite{adamic2008YA}; Baidu Knows
(Chinese)~\cite{jiang09baidu}; and Naver Knowledge iN
(Korean)~\cite{nam09naver}.  On each of the sites, millions of
questions are answered each month by individuals with a wide range of
expertise.

Online knowledge-sharing activity includes not just those who specialize in knowledge generation and dissemination, i.e. professional researchers and scholars, but also others who gained their expertise through study and experience. In addition to providing data on different types of
individuals, these datasets represent knowledge generation at
different scales. Authoring a research article or patent in
most cases involves weeks to years of research, culminating in a
significant new result worthy of publication. In contrast,
contributing a fact to Wikipedia or answering a question posed in an
online forum may involve little more than a simple recall of
previously attained knowledge -- and a few minutes of the contributor's
time.

In evaluating focus across such a broad range of activities, we aimed
to use a metric that captures three qualities: \emph{variety},
or how many different areas an individual contributes to;
\emph{balance}, or how evenly their efforts are distributed among
these areas; and \emph{similarity}, or how related those areas are~\cite{rafols08diversity}.  We use the
Stirling measure
$\mathcal{F}$, which captures all three aspects~\cite{stirling2007gfa}:
\begin{equation}
\mathcal{F} := \sum_{i,j} s_{i j} p_{i} p_{j}, \nonumber
\end{equation}
\noindent where $p_{i}$ is the proportion of the individual's
contributions in category $i$ and $s_{ij}=n_{ij}/n_{j}$ is a measure of similarity between categories $i$ and $j$, inferred from the number of joint contributors $n_{ij}$ between two categories $i$ and $j$. This $F$ metric assigns a narrower (higher) focus value to an individual who contributes to fewer,
related areas than to someone who contributes
in many unrelated areas.

The categories across which focus is measured differ by the type of knowledge-sharing medium.  An inventor's proportion $p_{c}$ of
contributions in subject $c$ is proportional to the number of times the class $c$ is assigned by the inventor or patent examiner to the inventor's patents. Articles in the APS dataset are classified according to the Physics and Astronomy Classification Scheme. For JSTOR articles, in the absence of a pre-defined category structure, we used unsupervised topic models on the full text of
authors' research articles~\cite{blei2003latent}.  Wikipedia articles are situated within Wikipedia's category hierarchy, while
answers provided in Q\&A forums are sorted according to the hierarchy of categories of
the corresponding questions.

For each dataset, we sought a relevant, objective measure of
quality of a contribution and evaluated it in the context of its peers.  For research
articles, we measured each article's citation count relative to those
of other articles in the same discipline and year~\cite{valderas2007}.
Likewise, patents' citation counts were compared with those of other
patents in the same patent classes and years. In doing this, we
control for discipline-specific factors that can impact a publication's citation count such as publication cycle length and number of publications in the discipline~\cite{stringer2008ejr,seglen1997ifj}.
For Wikipedia contributions, we consider the percentage of words an
author newly introduces to an article that survive subsequent
revisions~\cite{adler2007cdr}. 
Finally, for Q\&A forums, we rely on the asker's rating of answers: a good
contributor should have their answer selected as best more often than
expected by chance.

\begin{figure}[!ht]
\begin{center}
\includegraphics[width=0.55\textwidth]{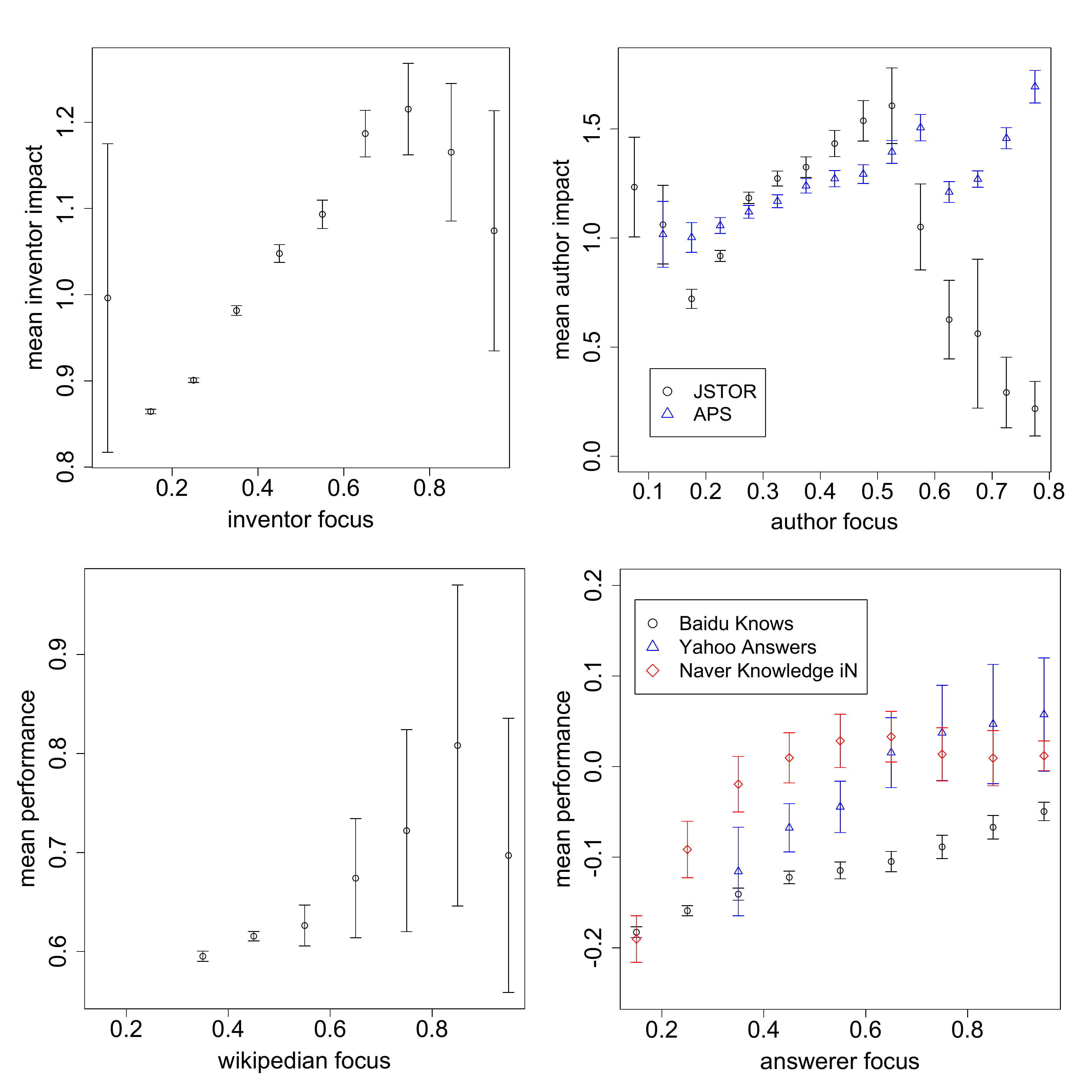} 
\end{center}
\caption{\textbf{Mean quality as a function of focus of contribution}.}
\label{fig:qvsf}
\end{figure}

\begin{figure}[!ht]
\begin{center}
\includegraphics[width=0.55\textwidth]{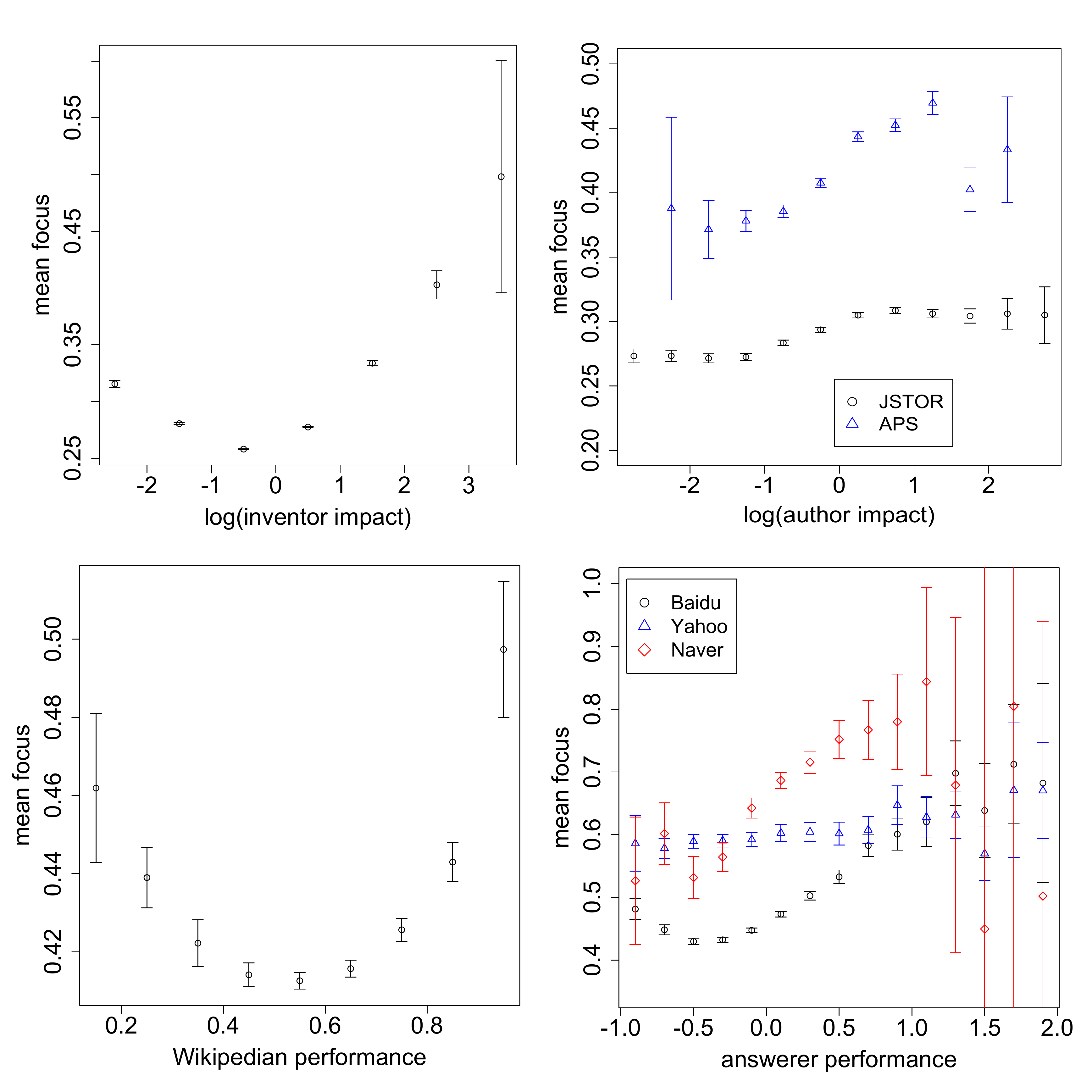} 
\end{center}
\caption{\textbf{Mean focus as a function of quality of contribution}.}
\label{fig:fvsq}
\end{figure}

Using these measures of focus and quality, we find that focus is
weakly yet consistently positively correlated with quality across all types of
knowledge contribution systems, as summarized in
Table~\ref{tab:focusQQ}. The relationship between focus and quality is detailed further in Figure~\ref{fig:qvsf}, which shows the variation in
average quality for individuals grouped by their levels of focus. As
focus increases, so does average quality; but the trend levels off or
even reverses for extremely focused individuals.  This is clarified by
plotting average focus at a given level of quality (see
Figure~\ref{fig:fvsq}). While high-quality contributors are more
narrowly focused than others on average, very poor contributors sometimes
also dwell in a single area. In Q\&A forums, we find further that
narrowly focused users with poor track records of giving best answers
tend to give answers that are significantly shorter than those of
other users.

\begin{table*}[!htpb]
\begin{center}
\caption{\bf{Pearson correlation between quantity, focus, and quality.}  All correlations are significant at 
$p < 0.001$. } \label{tab:focusQQ}
\begin{tabular}{l|l|l|l}\\
type & $\rho$(log(quantity),focus) & $\rho$(log(quantity),quality) & $\rho$(focus,quality) \\ \hline \hline
Research articles JSTOR &  -0.055 &  0.058 & 0.112\\ \hline
Research articles APS &  0.302 &  0.130 & 0.173\\ \hline
Patents & 0.286 & 0.100 & 0.094\\ \hline
Wikipedia & -0.274 & 0.133 & 0.084\\ \hline
Yahoo Answers & 0.150 & 0.116 & 0.084\\
Baidu Knows &0.095 & 0.083 &0.111\\
Naver KnowledgeIn &0.066 &0.102 &0.169\\
\hline \hline
\end{tabular}
\end{center}
\end{table*}

We note that these datasets provide incomplete views of contributors'
activity. JSTOR archives over a thousand journals, but does not include many more. Inventors' patents prior to 1976 are not captured in our data.  Likewise, we parsed
only a subset of the Wikipedia revision history; and while our Yahoo
Answers data set spanned the complete activity of a sample of users,
our Baidu and Naver data sets covered only several months each (see
supplementary materials). 

Nevertheless, we believe our results to be robust. We expect that we would find equally strong or stronger correlation between focus and quality, if we had complete records of each individual's contributions. An indication of this is that the correlations between focus and quality strengthen for users for whom we observe a higher number of contributions. For example, the correlation between focus and impact
for inventors with 10-20 patents is just $0.079\pm0.003$, but for
inventors with 50 to 100 patents it increases to
$0.166\pm0.014$. Similarly, Wikipedia users editing between 10 and 20
pages display a correlation of $0.088\pm0.065$, while those editing
between 50 and 100 pages display a correlation of $0.243\pm0.038$.

We find our results to be robust in several other respects as well.
Aside from being consistent across a wide range of media and
performance metrics, our results hold when focus is measured at different levels of granularity, e.g. when using top-level patent, Wikipedia, and Q\&A categories as opposed to subcategories in those data sets, and when we
construct 250 as opposed to 100 topics in the JSTOR dataset. While the
distribution of focus shifts downward as we increase the granularity,
the correlations between focus and other variables remain
qualitatively similar.

\begin{table*}[!htpb]
\begin{center}
\caption{\bf{Pearson correlations between quality, focus, and quantity,
  when self-citations are removed.} All correlations are significant at 
$p < 0.001$. } \label{tab:focusQQnoself}
\begin{tabular}{l|l|l|l}\\
type &$\rho$(log(quantity),quality) & $\rho$(focus,quality) \\ \hline \hline
research articles JSTOR & 0.051 & 0.095\\ \hline
research articles APS &    0.059 & 0.036\\ \hline
patents & 0.059 & 0.073 \\ \hline
\end{tabular}
\end{center}
\end{table*}

\begin{figure}[!ht]
\begin{center}
\includegraphics[width=0.55\textwidth]{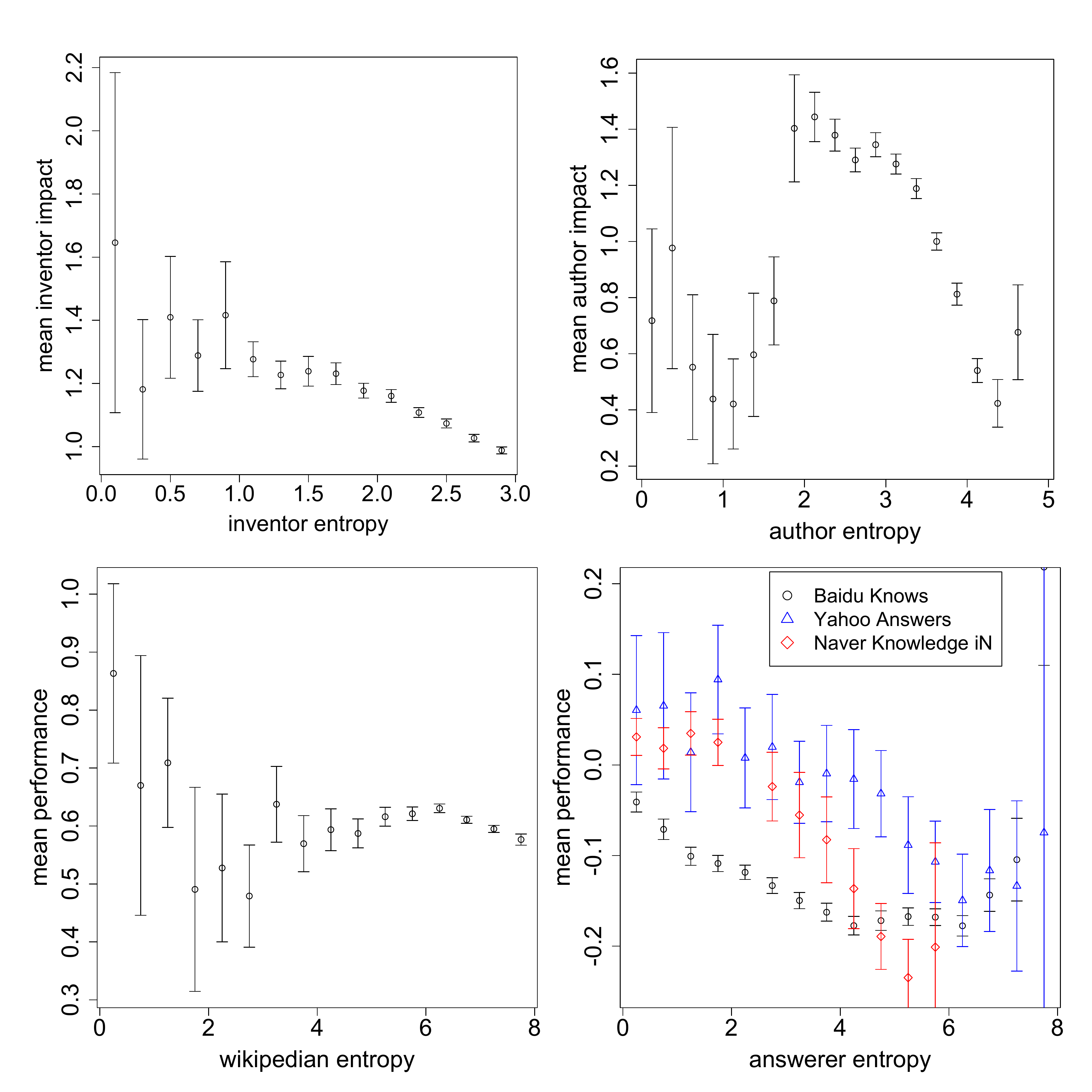} 
\end{center}
\caption{\textbf{Mean quality as a function of entropy of contribution.}}
 \label{fig:QvsE}
\end{figure}

\begin{figure}[!ht]
\begin{center}
\includegraphics[width=0.55\textwidth]{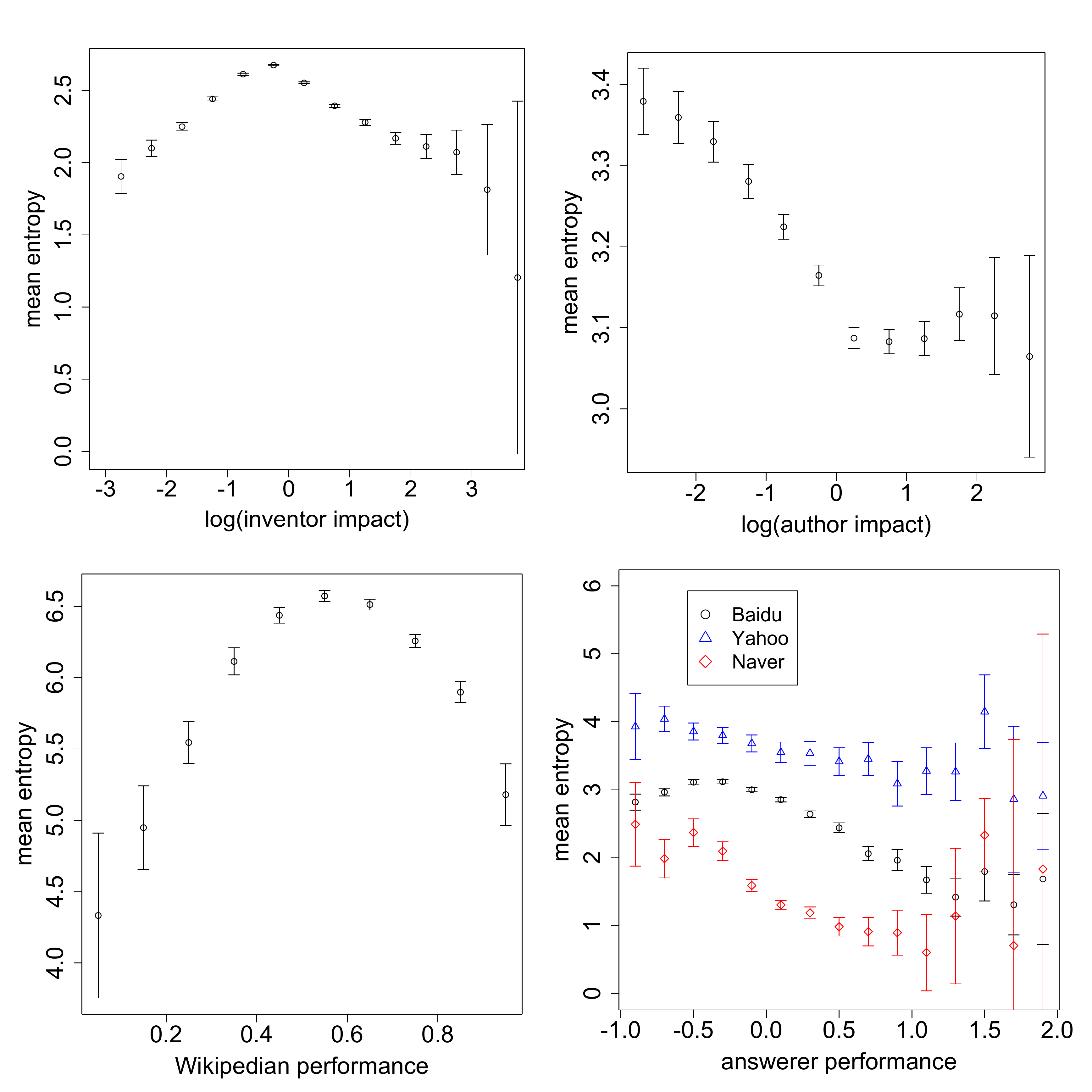} 
\end{center}
\caption{\textbf{Mean entropy as a function of quality of contribution}.}
 \label{fig:EvsQ}
\end{figure}

We also find our results to be consistent,
though weaker for the paper and patent data sets, when self-citations, i.e. citations between two papers that share an author with the same last name,  are removed. Table~\ref{tab:focusQQnoself} summarizes these findings. Self-citations may inflate the impact of
prolific and focused authors who have greater opportunity and
justification to cite their own work. 
We also find
consistent results using alternative focus measures such as Shannon
Entropy, $-\sum_{i} p_{i} ln(p_{i})$. Entropy captures the balance and variety of contribution, but
not similarity, and is negatively correlated with focus. We find the
results to be qualitatively consistent to those obtained using the
Stirling measure. Table~\ref{tab:entropyQQ} and Figures~\ref{fig:QvsE} and \ref{fig:EvsQ} correspond to Table~\ref{tab:focusQQ} and
Figures~\ref{fig:qvsf} and \ref{fig:fvsq} respectively, but present results using contributor entropy rather than focus. 

\begin{table*}[!htpb]
\caption{\bf{Person correlations between focus, entropy, quantity, and quality. } All correlations are significant at $p <
  0.001$.} \label{tab:entropyQQ}
\begin{tabular}{l|c|c|c}\\
data set &  $\rho$(entropy,focus) & $\rho$(log(quantity),entropy)  & $\rho$(entropy,quality) \\ \hline \hline
research articles JSTOR &  -0.606 &  0.082 & -0.218 \\ \hline
research articles APS &  -0.601 &  0.083 & -0.116 \\ \hline
patents & -0.848 & 0.160 &-0.117\\ \hline
Wikipedia & -0.755 & 0.629 & -0.103\\ \hline
Yahoo Answers & -0.878 & -0.081 & -0.121\\
Baidu Knows & -0.918 & -0.034 &-0.108\\
Naver KnowledgeIn & -0.946 &-0.175 &-0.223\\
\hline \hline
\end{tabular}
\end{table*}

One remaining concern is that focus and quality are both correlated with a third variable that holds greater explanatory power. One such potential variable is that of quantity. Quantity itself is positively correlated with quality, revealing a possible link between contributor success and motivation or
resources. However, focus remains a significant factor in
the quality of contributions, even once quantity is accounted for (see Table~\ref{tab:regression}). We also note that quantity's correlation with focus varies by medium studied, as shown in Table~\ref{tab:focusQQ}.
The correlation is positive for patents
and Q\&A forums, but negative for research and Wikipedia
articles. Individuals generating many patents or answers tend to focus
their contributions more narrowly, but authors who write a greater
number of research and Wikipedia articles tend to make broader contributions. 

\begin{table*}[!htpb]
\caption{\bf{Regression models using quantity and focus as predictors of quality}} \label{tab:regression}
\begin{tabular}{cccccccc}\\
variable & JSTOR & APS & patents & Wikipedia & YA & Baidu & Naver \\ \hline \hline
focus &  1.899$^{***}$ & 0.553$^{***}$ & 0.501$^{***}$ & 0.353$^{***}$  & 0.245$^{***}$ & 0.143$^{***}$ & 0.146$^{***}$ \\
	& (0.085) & (0.040) & (0.009) & (0.033)  & (0.046) & (0.005) & (0.015)\\
log(quantity) & 0.177$^{***}$ & 0.124$^{***}$ & 0.163$^{***}$ & $0.027^{***}$& 0.046$^{***}$ & 0.052$^{***}$ & 0.039$^{***}$ \\
& (0.014) & (0.008) & (0.003) & (0.002) & (0.005) & (0.002) & (0.008) \\
\hline
$R^{2}$ &0.017 & 0.025 & 0.015 & 0.023 & 0.022 & 0.023 & 0.022\\ \hline \hline
\end{tabular}
\end{table*}

Finally we examine whether individuals narrow or broaden their focus over time.  For JSTOR research articles,
patents, and Q\&A forums, a majority of contributors narrow their focus over time (see
Table~\ref{tab:focuschange}). Wikipedia contributors and physicists,
on the other hand, do not appear to specialize further. In addition to
a change in focus, we also observe a slight change in quality. Across
most datasets, contributors tend to improve the quality over time; exceptions include Baidu
Knows, where the change in answer quality is not significant, and JSTOR, where there is a
statistically significant decline in contribution quality.  One might speculate that a researcher's early success permits him or her to continue producing publications, but that the quality of those publications may fall due to factors such as moving from a primary contributor to a project management role.

\begin{table*}[!htpb]
\begin{center}
\caption{\bf{Change in focus from first half to second half of contributions.}}
\label{tab:focuschange}
\begin{tabular}{l|c|c|c|}  \\
data set & \% who increased focus & av. change focus & av. change quality \\ \hline \hline
research articles JSTOR & 62.0\% &  0.024 & -0.394 \\ \hline
research articles APS & 44.0\% &  -0.012 & -0.157 \\ \hline
patents & 79.5\% &  0.139  & 0.080  \\ \hline
Wikipedia &  49.5\% &   not. sig.  & 0.066 \\ \hline
Yahoo Answers &  69.8\% &  0.056 &  0.056 \\
Baidu Knows & 61.4\% & 0.040   &  not. sig. \\
Naver KnowledgeIn &    68.8\%& 0.023  &  0.034 \\
\hline \hline
\end{tabular}
\end{center}
\end{table*}

\section*{Discussion}
In conclusion, we have quantified the value an individual's
focus in contributing knowledge through different media and across a
wide range of fields. Consistently we observe a slight but
significant correlation between an individual's degree of focus and
this individual's quality of contribution. The relationship persists
even when quantity of contributions is taken into account. 

How should individuals invest their time? While our results do not demonstrate
causality, the overall trend appears to favor those who do a few
things and do them well. However, individuals who focus in a very
narrow field tend to contribute work that is on average less-well
recognized than that of their slightly less focused peers.

This work immediately suggests several areas for future research.  It
would be useful to understand the benefit of narrowing one's focus in
the context of the research field and media type, in addition to the
quality and diversity of one's prior efforts. One could also examine whether highly focused authors of research articles or patents may be introducing marginal or
incremental contributions.

In addition, while these results have shed light on the value of focus
in the context of the individual, they say nothing about focus
  in the context of a group.  After all, several studies have
demonstrated the value of interdisciplinary collaborations in the
sciences, and we believe that large-scale online knowledge sharing
systems such as those discussed are successful precisely because they
bring together individuals of different backgrounds. This leaves open
the question of whether interdisciplinary collaborations between more
{\em focused}--yet \emph{collectively diverse}--individuals are more
fruitful.

\section*{Materials and Methods}
Table~\ref{tab:datalist} summarizes the seven data sets we used to study
focus and contribution. For each dataset we selected a threshold criterion for
the minimum level of activity needed for an individual to be included.

\begin{table*}[!htpb]
\begin{center}
\caption{\bf{Description of data.}}\label{tab:datalist}
\begin{tabular}{l|rrl} \hline
dataset & time span & \# individuals & threshold \# contributions\\ \hline \hline
JSTOR & 1668-2006 & 37,031 &  10\ articles\\ 
APS & 1977-2006 & 22,351 &  10\ articles\\ \hline
patents & 1976-2006 & 604,113 & 10 patents\\ \hline
Wikipedia & 2001-2006 & 7,129 & 40 edits\\ \hline
Yahoo Answers & 08/05-03/09 & 5,256 & 40 answers\\
Baidu Knows & 12/07-05/08 & 65,854 & 40 answers\\
Naver KnowledgeIn & 12/08-02/09 & 5,918 & 40 answers\\
\hline \hline
\end{tabular}
\end{center}
\end{table*}

\textbf{Research articles}. A snapshot of \textbf{JSTOR} data includes 2.0
million research articles with 6.6 million citations between them.
JSTOR spans over a century of publications in the natural and social
sciences, the arts, and humanities. For this data set, we needed to address name ambiguity.
For example there were 26,000 instances where a person with the last
name of Smith authored an article and 728 unique combinations of
initials appearing alongside ``Smith''. Identifying two different
individuals as being one and the same would tend to introduce data
points with low focus and an inflated number of articles. Since both
variables are related to quality, we sought to exclude such
instances. We excluded authors with $\sqrt{F_{L}*L_{F}} > 200$, where
$F_{L}$ is the number of first names or initials the inventor's last
name occurs with in the data set, and $L_{F}$ is the number of last
names the inventor's first name occurs with. We also collapsed matching names and
initials if there was only one matching first name/inital pair and
the last name occurred with fewer than 50 first names. This left us
with 37,031 authors with 10 or more publications, for whom we were
reasonably sure that they were uniquely identified.

Using Latent Dirichlet Allocation~\cite{blei2003latent}, we generate 100
topics over the entire corpus of research articles. Each document was
assigned a normalized score for each of the 100 topics, and the
pairwise topic similarity matrix $s$ was computed from cosines of
vector values across documents. An author's distribution across topics
was computed by averaging the topic vectors of all of the articles they
authored. For robustness, we repeated the analysis with 250 topics
instead of 100, and found quantitatively similar correlation between
focus and quality, although focus scores were lower due to the finer
granularity. The quality of an article is measured as the number
  of times the article is cited, divided by the number of times other
  articles in the same area and year are cited. Citations originate
  within the dataset. By normalizing quality by area, we mitigate the
  possible biases introduced by some areas being better represented in
  the dataset than others.

Our database of \textbf{American Physical Society} publications included
Physical Review Letters, and Physical Review A-E journal articles. We
excluded Reviews of Modern Physics as we were considering the impact
of original research rather than review articles. The data set
contained 396,134 articles published between 1893 and 2006, with
3,112,706 citations between them. For our purposes, we were limited to
the 261,161 articles with PACS (Physics and Astronomy Classification
Scheme) codes associated with articles published after 1977. The PACS
hierarchy has 5 levels, and we performed our analysis at the level of
the 76 main categories, such as 42 (Optics) and the 859 2nd level
categories, e.g. 42.50 (Quantum Optics).

\textbf{ Patents}. The patent data set contains all 5,529,055 patents
filed between 1976 and 2006, in 468 top level categories. We construct
a similarity matrix for the 468 categories, reflecting the frequency
with which inventors in one category also file patents in
another. There are 3,643,520 patents citing 2,382,334 others, for a
total of 44,556,087 citations. We excluded inventors with
$\sqrt{F_{L}*L_{F}} > 150$.  This makes it unlikely that we would
identify two separate individuals as being one. We measure an
inventor's impact according to a citation count normalized by the
average number of citations for other patents in the same year and
categories as those filed by the inventor.

\textbf {Q\&A forums}: We obtained snapshots of activity on Q\&A
forums with uniquely identified users posting answers to questions in
distinct categories. We perform our analysis at the subcategory level,
which gives us enough resolution to differentiate the question topics,
while supplying a sufficient number of observations in each
subcategory. We use best answers as a proxy for answer quality. The
best answer is selected by the user who posed the question. If this
user does not select a best answer, it may be selected via a vote by
other users. The quality metric we used was the $\gamma$ score, which
compares the number of answers the user gives that were selected as
best among others, relative to the expected number of best
answers. The expected number depends on the number of other answers
provided to each question. $\gamma = $ {\em (observed -
  $b_{e})/b_{e}$}. $b_{e}$ the expected number of best answers is
simply given by $b_{e} = \sum_{k}1/a_{k}$, where $a_{k}$ is the total
number of users answering question $k$.

\textbf {Wikipedia}: Our Wikipedia dataset is a meta-history dump file
of the English Wikipedia generated on Nov. 4th, 2006. The
dump file has the entire revision history of about 1.5 million
encyclopedia pages, of which we parsed 100,000, or about
7\%. In order to verify that our sample is unbiased with
  respect to topic distribution, we compare the category and
  subcategory distributions of our sample to that of a larger corpus
of 1 million pages. The two distributions have a nearly
perfect correlation ($\rho > 0.96^{***}$).

Articles are a product of varying number of revisions, from several to
10,000 for single article. Revisions are contributed
by either registered or anonymous users. Since anonymous users'
revision histories are non-traceable, we only consider registered users
whose unique user names are associated with at least 40
revisions. We excluded Wikipedia administrators from our study because
they may perform a primarily editorial role. In like manner, to better filter the noise of measuring the
quality of words by the final version of articles, we only choose
pages in which ewer than 5\% of the revisions occurred in the
  30 days prior to the data dump.

A Wikipedia contributor's focus and entropy were calculated from the
second-level categories of the pages they edit. Each Wikipedia article
belongs to one or more categories. We truncated each hierarchical category 
to one of the roughly 500 second-level categories.

The quality of a contribution is measured in terms of $w_{new}$, the
number of new words added by a user to Wikipedia articles, such that
the words were not present in any previous revisions of those
articles.  We found a high correlation between the number of new words
that survive 5 revisions, and the number $w_{surv}$ that survive to
the last revision of the article ($\rho > 0.97^{***}$), consistent
with previous analyses of edit persistence~\cite{panciera09wiki}. We
therefore constructed a simple metric by taking the proportion of new
words introduced by the user that are retained in the last version of
a sufficiently frequently edited article:
$w_{surv}/w_{new}$~\cite{adler2007cdr}.

\section*{Acknowledgments}
We thank IBM for providing the patent data, and JSTOR, APS, and Katy Borner
  for providing the article citation data. We would also like to thank
  Michael McQuaid, Jure Leskovec, Scott Page and Eytan Adar for helpful comments. 
 This research was supported by MURI award FA9550-08-1-0265 from the Air Force
 Office of Scientific Research  and NSF award IIS 0855352.

\bibliographystyle{plain}

\end{document}